\documentclass[11pt]{article}
\usepackage{fullpage}
\usepackage{graphicx}
\usepackage{amssymb}
\usepackage{amsmath}
\usepackage{theorem}
\usepackage{enumerate}
\usepackage{filecontents}
\usepackage{csvsimple}
\usepackage{natbib}
\usepackage{bibentry}
\usepackage{color}
\usepackage{colortbl}
\usepackage{ulem}
\usepackage{tcolorbox}
\usepackage{graphicx}
\usepackage{tikz}
\usepackage{xcolor}
\usetikzlibrary{fit,positioning}
\usepackage{booktabs}

\usepackage{rotating}

\usepackage[affil-it]{authblk}

\usepackage{varwidth}

\nobibliography*

\date{}

\title{Loss-function learning for digital tissue deconvolution}
 \author{Franziska G\"ortler$^1$, Stefan Solbrig$^2$, Tilo Wettig$^2$, Peter J. Oefner$^3$, \\Rainer Spang$^1$, Michael Altenbuchinger$^1$\footnote{michael.altenbuchinger@ukr.de}}
\affil{$^1$Statistical Bioinformatics, Institute of Functional Genomics, University of Regensburg, Am Biopark 9, 93053 Regensburg, Germany \\
$^2$Department of Physics, University of Regensburg, Universit\"atsstra\ss e 31, 93053 Regensburg, Germany\\ 
$^3$Institute of Functional Genomics, University of Regensburg, Am Biopark 9, 93053 Regensburg, Germany 
}
                     
\begin{document}
\maketitle
\begin{abstract}
\noindent\textbf{Background:} The gene expression profile of a tissue averages the expression profiles of all cells in this tissue. Digital tissue deconvolution (DTD) addresses the following inverse problem: Given the expression profile $y$ of a tissue, what is the cellular composition $c$ of that tissue?
If $X$ is a matrix whose columns are reference profiles of individual cell types, the composition $c$ can be computed by minimizing $\mathcal L(y-Xc)$ for a given loss function $\mathcal L$. Current methods use predefined all-purpose loss functions. They successfully quantify the dominating cells of a tissue, while often falling short in detecting small cell populations. \\

\noindent\textbf{Results:} Here we learn the loss function $\mathcal L$ along with the composition $c$. This allows us to adapt to application-specific requirements such as focusing on small cell populations or distinguishing phenotypically similar cell populations. Our method quantifies large cell fractions as
accurately as existing methods and significantly improves the detection of small cell populations and the distinction of similar cell types.
\end{abstract}

\section{Introduction}
Different tissues of the body have different cellular compositions. The composition of tumor tissue is different from that of normal tissue. 
Also, when comparing two tumor tissues, their cellular composition can differ greatly. 
The relatively small populations of tumor-infiltrating immune cells are of particular importance. 
They affect progression of disease  \citep{galon2006type} and success of treatment \citep{fridman2012immune}. Immune therapies block communication lines between tumor cells and 
infiltrating immune cells. Whether they are successful or not depends on the presence, quantity, and molecular sub-type of the infiltrating 
immune cells \citep{hackl2016computational}. Immune-cell populations are typically small, and their molecular phenotype can be difficult to observe under the microscope.
Single-cell technologies such as fluorescence-activated cell sorting (FACS; \cite{ibrahim2007flow}), cytometry by time-of-flight 
(CyTOF; e.g. \cite{bendall2011single}), and single-cell RNA sequencing \citep{wu2014quantitative} assess molecular features on the single-cell level and can thus 
be used to determine the cellular tissue composition experimentally. 

A more cost- and work-efficient alternative to single-cell assays is a combination of bulk-tissue gene expression profiling with 
digital tissue deconvolution (DTD) \citep{lu2003expression, abbas2009deconvolution, gong2011optimal, qiao2012pert, altboum2014digital, newman2015robust, li2016comprehensive}.
DTD addresses the following inverse problem: given the bulk gene expression profile $y$ of a tissue, what is the cellular composition $c$ of 
that tissue? 
Supervised DTD assumes that there is a matrix $X$ whose columns are reference profiles of individual cell types. 
The composition $c$ of $y$ can be computed by minimizing $\mathcal L(y-Xc)$ for a given loss function $\mathcal L$. 
Competing DTD methods use different predefined all-purpose loss functions $\mathcal L$ and different estimation algorithms 
to distil $c$ from $y$ and $X$. 

The practical objective of DTD is to estimate $c$ correctly, while the formal objective of common DTD algorithms is to estimate $y$ correctly. If tissue expression profiles were exact mixtures of reference profiles, existing methods should work perfectly. They are not and this causes problems:

\begin{itemize}

\item[(1)]{\textbf{Collections of references profiles can be incomplete.}
There might be cells in the tissue that are not represented by the reference profiles. In that case the global DTD problem is not solvable, and DTD-algorithms will compensate for the contributions of these cells by increasing the frequencies of other cell types.}

\item[(2)]{\textbf{Small cell fractions are hard to quantify.}
From a practical point of view this is probably the most important point, and improvements are needed badly. Immunological cell populations in a tumor are small, 
but they may determine the reaction of a tumor to immunotherapy. Therefore, DTD algorithms must use faint signals from small cell populations more effectively.}

\item[(3)]{\textbf{Some cell types can hardly be distinguished by their expression profiles.}
The profile of an epithelial cell differs greatly from that of a lymphoid cell. For two immunological sub-entities of CD8+ T cells the differences are more subtle. The more similar two cell types are, the more similar are their expression profiles, and the more difficult is their distinction.}
\end{itemize}

In summary, different applications need different approaches. One way to adapt the estimation of $c$ is to adapt the loss function  $\mathcal L$.
If the focus of an application is on a predefined set of cell types, genes that are informative to distinguish exactly these cells should dominate $\mathcal L$. This
is even more important if the focus is on small cell populations, the faint signals of which must not be suppressed. 
Unfortunately, it is not clear \textit{a priori} which genes to ignore and which to focus on.

\section{Methods}

\subsection{Notations}
Let $X\in  \mathbb{R}^{p \times q}$ be a matrix with cellular reference profiles $X_{\cdot, j}$ in its columns, where the dot stands for all row indices. 
$X_{ij}$ is the reference expression value of gene $i$ in cells of type $j$, 
$p$ the number of genes, and $q$ the number of cell types in $X$, respectively. 
We further introduce a matrix $Y\in \mathbb{R}^{p\times n}$ with bulk profiles of $n$ cell mixtures 
$Y_{\cdot, k}$ in its columns and a matrix $C\in \mathbb{R}^{q\times n}$ with the cellular compositions 
of the mixtures $C_{\cdot, k}$ as columns.

\subsection{Loss-function learning}
Following the established linear DTD algorithms, we approximate the mixture $Y_{\cdot,k}$ by a linear 
combination of reference profiles (the columns of $X$) with $C_{.,k}$ as weights and estimate the 
composition of the $k$-th mixture $C_{\cdot,k}$ by minimizing 
\begin{equation}
\mathcal L_g(Y_{\cdot,k}-XC_{\cdot,k})\,,
\end{equation}
where 
\begin{equation} 
\mathcal L_g= || \textrm{diag}({g}) (Y_{\cdot,k}-XC_{\cdot,k}) ||^2_2 \,.
\label{L}
\end{equation}
In contrast to standard DTD algorithms, which determine $g$ by prior knowledge or separate statistical analysis, 
we will learn $g$ directly from data.
To this end we assume that we have a training set of mixtures $Y_{\cdot,k}$ 
from a specific application context with known cellular compositions 
$C_{\cdot,k}$.  
The entries of ${g}$ are the gene weights that define the 
loss function.
We want to learn $g$ from the training data such that minimizing 
$\mathcal L_{g}(y-Xc)\,$ with respect to $c$
yields accurate quantifications of cell populations for future samples with 
similar characteristics as those used for training.  

Our method has two nested objective functions: An outer function $ L(g)$ and 
an inner function $\mathcal L_g$, which is here given by equation (\ref{L}). 
$ L$ evaluates discrepancies 
between the estimated and the true cellular frequencies of cell types across samples:
\begin{equation} 
 L({g}) =-\sum_{j=1}^q\textrm{cor}(C_{j,\cdot},\hat C_{j,\cdot}({g}))
\quad\mbox{subject to } g_i\ge0\, \mbox{and } ||g||_2=1\,,
\label{L_out}
\end{equation}
where the $\hat C_{j,\cdot}({g})$ are the estimates of $C_{j,\cdot}$ given $g$.
To evaluate $L( g)$  we need to calculate all $\hat C_{j,\cdot}({g})$, which requires 
optimizing $\mathcal L_g$ with respect to all $C_{\cdot, k}$.
Note that if  $\hat g$ is a minimum of $ L$, so is $\alpha \hat {g}$. The
constraint $|| g||_2=1$ is thus needed to ensure unique solutions.

The minimum of $\mathcal L_g$ can be calculated analytically, yielding
\begin{equation}
\hat C({g}) = (X^T \Gamma X)^{-1} X^T \Gamma Y\,
\label{AnaSol}
\end{equation}
with $\Gamma=\textrm{diag}(g)$. Inserting this term into $L$ leaves us with a single optimization problem in $g$.
We minimize $L$  by a gradient-descent algorithm.
Let $\mu_j$ and $\sigma_j$ be the mean and standard deviation of $C_{j,\cdot}$, respectively. 
We obtain the gradient 
\begin{eqnarray}
\frac{\partial  L(g)}{\partial g_i}
    &=&\sum_{j=1}^q \sum_{k=1}^n \frac{1}{\sigma_j\hat \sigma_j}
	   \left(\frac{\textrm{cov}(C_{j,\cdot},\hat C_{j,\cdot})}{n\hat\sigma_j^2}
	   \big(\hat C_{jk} - \hat{\mu}_j\big)
	   - \frac{1}{n}(C_{jk}- \mu_j) \right)
	   \frac{\partial \hat C_{jk}(g)}{\partial g_i}\,
\label{gradient}
\end{eqnarray}			
with 
\begin{eqnarray}
  \frac{\partial \hat C(g)}{\partial g_i} = 
 (X^T\Gamma X)^{-1} X^T \delta(i) \big(1-X  (X^T \Gamma X)^{-1} X^T \Gamma \big)Y\,,
 %}
 \label{dCdg}
\end{eqnarray}
where $\delta(i)\in \mathbb{R}^{p\times p}$ is defined as
\begin{equation}
  \delta(i)_{jk}= \Big\{\begin{array}{cl} 1 &\mbox{if}\; i=j=k,\\ 0&\mbox{else.}\end{array}\,
\end{equation}
The constraints $||g||_2=1$ and $g_i\ge0$ were incorporated  by normalizing $g$ by its length and by restricting the search space to $g_i\ge0$.

\section{Results}
\subsection{DTD of melanomas}
 \label{data}
For both training and validation we need expression profiles of cellular mixtures of known composition. We used expression data
of melanomas whose composition has been experimentally resolved using single-cell RNAseq profiling \citep{tirosh2016dissecting}.
These data included 4,645 single-cell profiles from 19 melanomas. The cells were annotated as   
T cells (2,068), B cells (515), macrophages (126), endothelial cells (65), cancer-associated fibroblasts (CAFs) (61), 
natural killer (NK) cells (52), and tumor/unclassified (1,758). 
The first 9 melanomas  defined our validation cohort and the remaining 10 our training data. 

First, data were transformed into transcripts per million. Then, for each cell  cluster we sampled $20\%$ of single-cell profiles in the training data, summed them up,  normalized them
to a common number of counts, and removed them from the training data. This yielded reference profiles $X_{\cdot, j}$. The 1,000 genes with the highest variance across all reference profiles were used to train models. 

The sum of all single-cell profiles of a melanoma gave us bulk profiles. In addition, we generated a large number of artificial 
bulk profiles by randomly sampling single-cell profiles and summing them up. All bulk profiles 
were normalized to the same number of reads as those in $X_{\cdot, j}$.

\begin{figure}
\begin{minipage}{0.59\textwidth}
\centering
\includegraphics[width=1\textwidth]{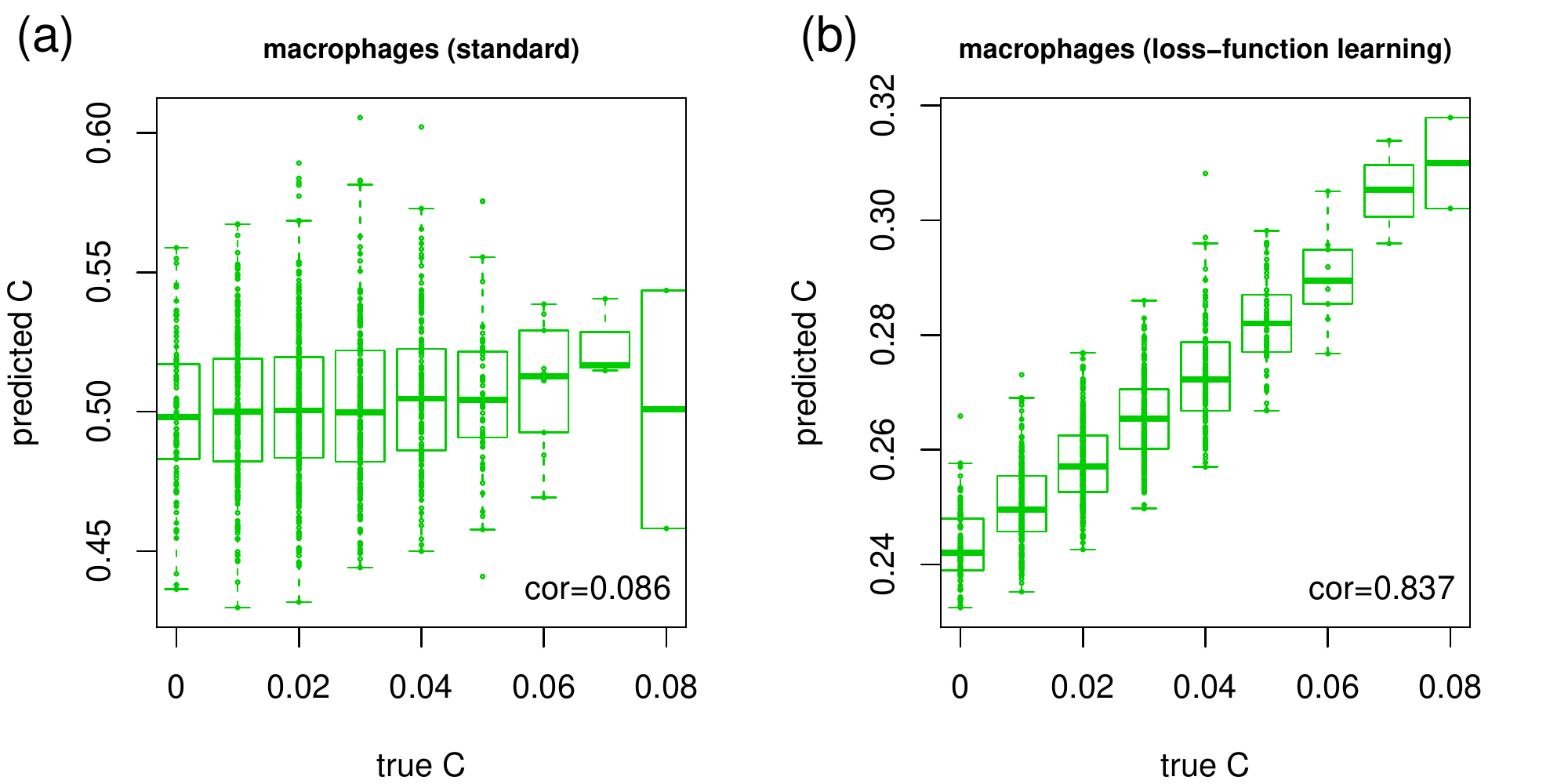}
\end{minipage}
\begin{minipage}{0.4\textwidth}
\caption{\label{M_simulation}Deconvolution performance with only a single reference profile (macrophages). Predicted cell frequencies are plotted versus real frequencies. Results from the standard DTD model with $g=1$ are shown in (a), for DTD with loss-function learning in (b). }
\end{minipage}
\end{figure}

\subsection{Loss-function learning improves DTD accuracy in the case of incomplete reference data}
\label{incomp}
We generated 2,000 artificial cellular mixtures from our training cohort. For each of these mixtures, we randomly drew 100 
single-cell profiles, summed up their raw counts, and normalized them to a fixed number of total counts. Analogously, we 
generated 1,000 artificial cellular validation mixtures. 

Then, we restricted $X$ to three cell types (T cells, B cells, and macrophages). Hence endothelial cells, CAFs, NK cells and tumor/unclassified cells in the mixtures are not represented in $X$. For standard DTD with $g=(1,\ldots,1)$, we observed correlation coefficients
of 0.70 (T cells), 0.39 (B cells), and 0.52 (macrophages) between true and estimated cell population sizes for the validation mixtures. These improved to 0.86 (T cells), 0.89 (B cells), and 0.83 (macrophages) for loss-function learning, after we ran 1000 iterations of the gradient descent algorithm on the training data.

To test the limits of the approach, we excluded all but the macrophages, which account for less than $3\%$ of all cells, from 
the reference data $X$. We observed, that standard DTD broke down, while loss-function learning yielded a model that predicted macrophage abundances 
that still correlated well ($r=0.84$) with the true abundances (Figure  \ref{M_simulation}).

\subsection{Loss-function learning improves the quantification of small cell populations}
\begin{figure}
\centering
\begin{minipage}{0.39\textwidth}
\includegraphics[width=1\textwidth]{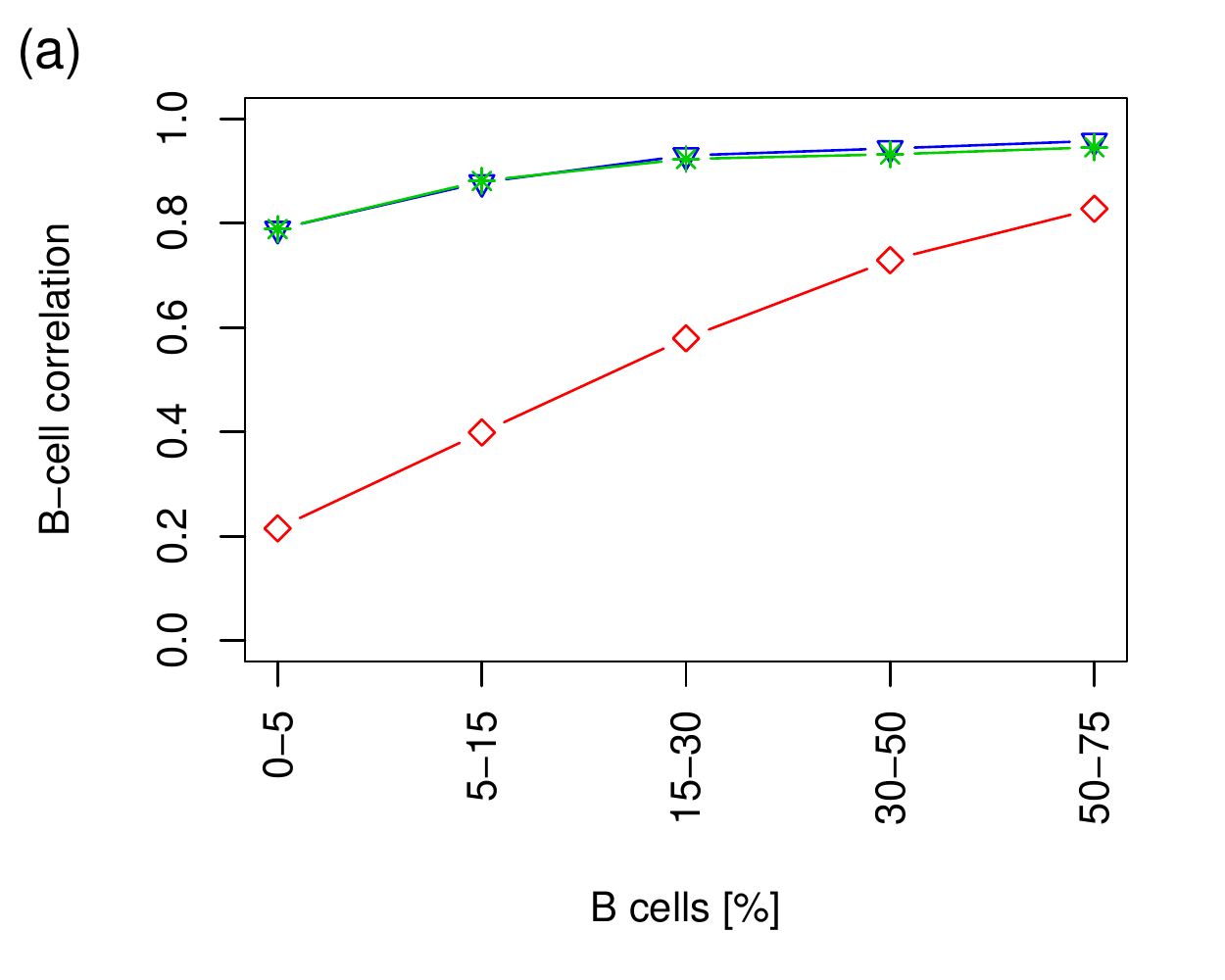}
\end{minipage}
\begin{minipage}{0.6\textwidth}
\includegraphics[width=1\textwidth]{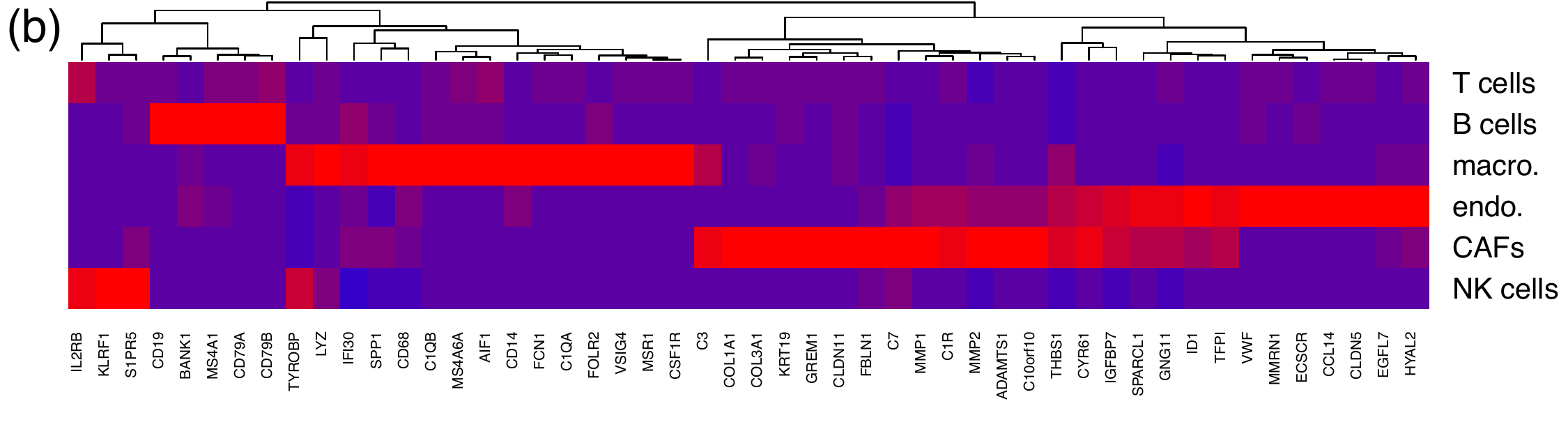}
\includegraphics[width=1\textwidth]{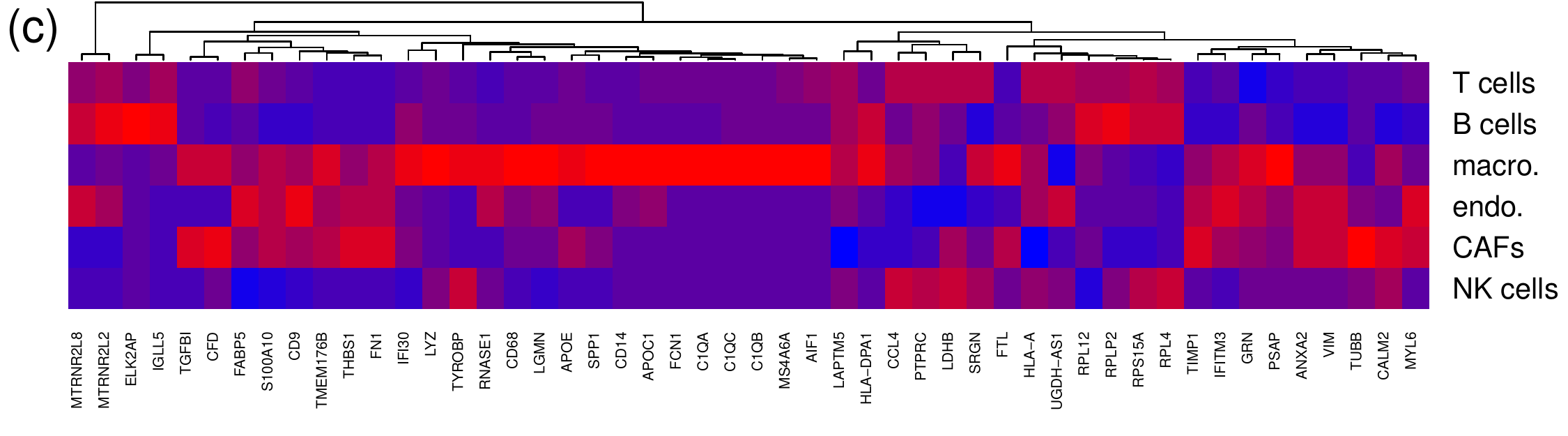}
\end{minipage}
\caption{\label{Sim_Num} 
Plot (a) shows how the correlation between predicted and true cellular frequencies for B cells depends on the proportion of B cells. The blue triangles correspond to models from loss-function learning and red diamonds to the standard DTD model with $g=1$. Furthermore, the green stars show how the model trained on mixtures with $0$ to $5\%$ B cells extrapolates to higher B-cell proportions.
 Plot (b) shows a heatmap of the 50 most important genes corresponding to the green star model (genes were ranked by $\hat g_i\times\textrm{var}(X_{i,\cdot})$). Plot (c) shows an analogous heatmap for loss-function learning on macrophages only. Blue corresponds to low expression and red to high expression.
  }
\end{figure}

We generated data as above but this time controlling the abundance of B cells in the simulated mixtures at
0 to 5 cells, 5 to 15, 15 to 30, 30 to 50, and 50 to 75 out of 100 cells. Not surprisingly, small fractions of B cells were harder to quantify than large ones. Loss-function learning
improved the accuracy for all amounts of B cells, but the improvements were greatest for small amounts (Figure \ref{Sim_Num}a). 
With only 0 to 5 cells in a mixture the accuracy improved from $r=0.22$ to $r=0.79$. Furthermore, we observed that loss-function learning on small B-cell proportions yielded a model that was highly predictive of B-cell contributions over the whole spectrum (Figure \ref{Sim_Num}a green stars). 

If we compare the top-ranked genes of the model learned for the small B-cell population (Figure \ref{Sim_Num}b) to that of the macrophage-focussed simulation (Figure \ref{Sim_Num}c), we observe that the former still comprises marker genes to distinguish all cell types, while the latter focusses on genes that characterize macrophages.

\subsection{Loss-function learning improves the distinction of closely related cell types}
\label{T_cell_subtypes}
The cell types that were annotated by \cite{tirosh2016dissecting} displayed very different expression profiles.
If we are interested in T-cell subtypes such as CD8+ T cells, CD4+ T-helper (Th) cells, and regulatory T cells (Tregs),
reference profiles are more similar and DTD is more challenging.
We subdivided the fraction of annotated T-cell profiles as follows: 
all T cells with positive CD8 (sum of CD8A and CD8B) and zero CD4 count were labelled CD8+ T cells (1,130).
Vice versa, T cells with zero CD8 and positive CD4 count were labelled CD4+ T cells (527).
These were further split into Tregs if both their FOXP3 and CD25 (IL2RA) count was positive (64),
and CD4+ Th cells otherwise (463).  
T cells that fulfilled neither the CD4+ nor the CD8+ criteria (411) contributed to the mixtures, but were not assessed by DTD.
We augmented the reference matrix $X$, here consisting of T cells, B cells, macrophages, endothelial cells, CAFs and NK cells, by these cell types, replacing the original all T-cell profile with the more specific profiles for CD8+, CD4+ Th and Tregs. Then we simulated 2,000 training and 1,000 test mixtures as described above.

\begin{figure}
\begin{minipage}{0.57\textwidth}
\centering
\includegraphics[width=1\textwidth]{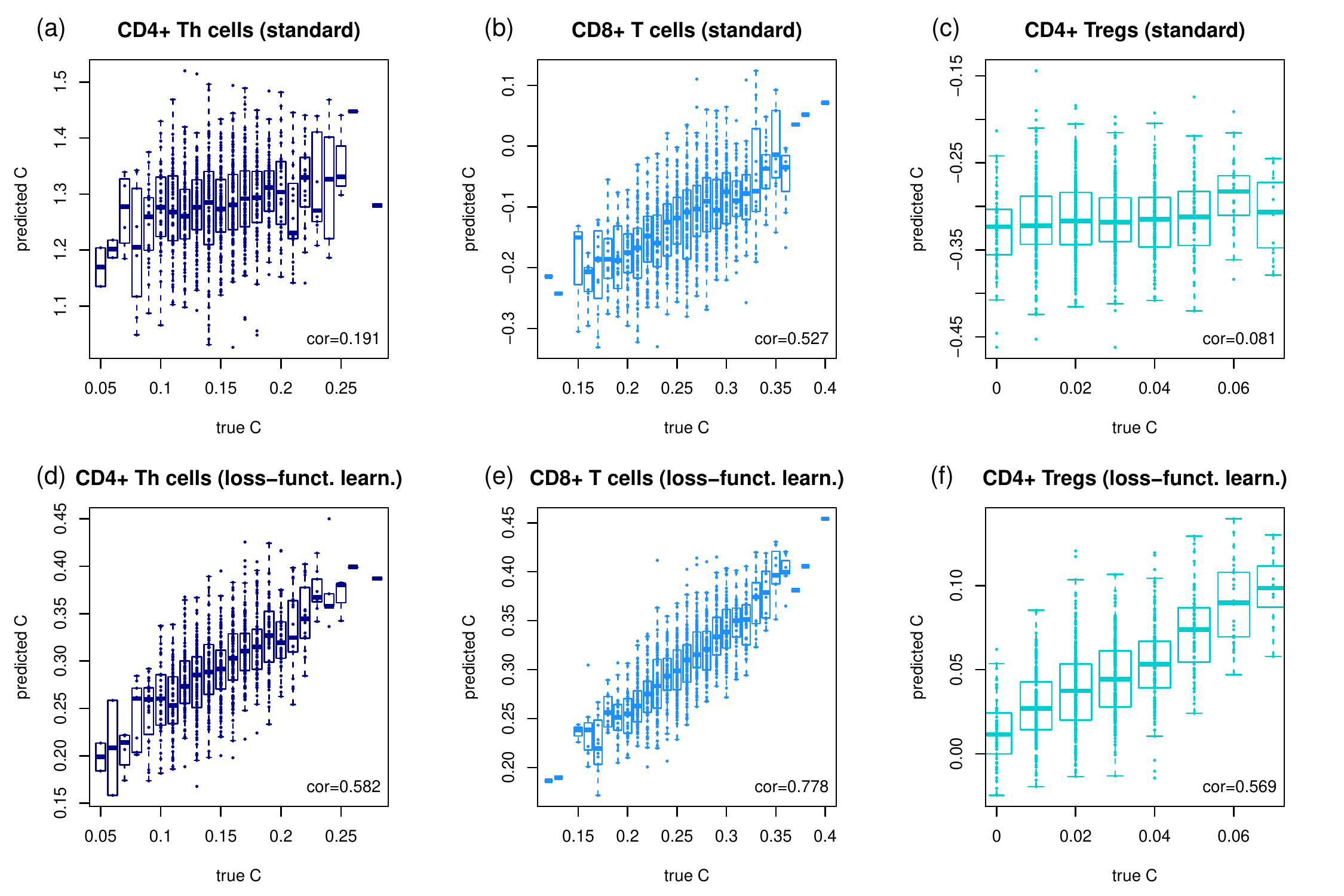}
\end{minipage}
\begin{minipage}{0.42\textwidth}
\caption{\label{Tcells}Deconvolution of T-cell subentities. Results from the standard DTD model with $g=1$ are shown in the upper row, plots (a-c), results from loss-function learning in the lower row, plots (d-e). }
\end{minipage}
\end{figure}

For standard DTD with $g=1$ we observed correlation coefficients
of 0.19 (CD4+ Th), 0.53 (CD8+), and 0.08 (Tregs) between true and estimated cell population sizes.
These improved to 0.58 (CD4+ Th), 0.78 (CD8+), and 0.57 (Tregs) for our method 
(Figure  \ref{Tcells}).

\subsection{Loss-function learning is beneficial even for small training sets, and the performance improves as the training dataset grows}
\label{TrSize}
\begin{figure}
\begin{minipage}{0.35\textwidth}
\centering
\includegraphics[width=1\textwidth]{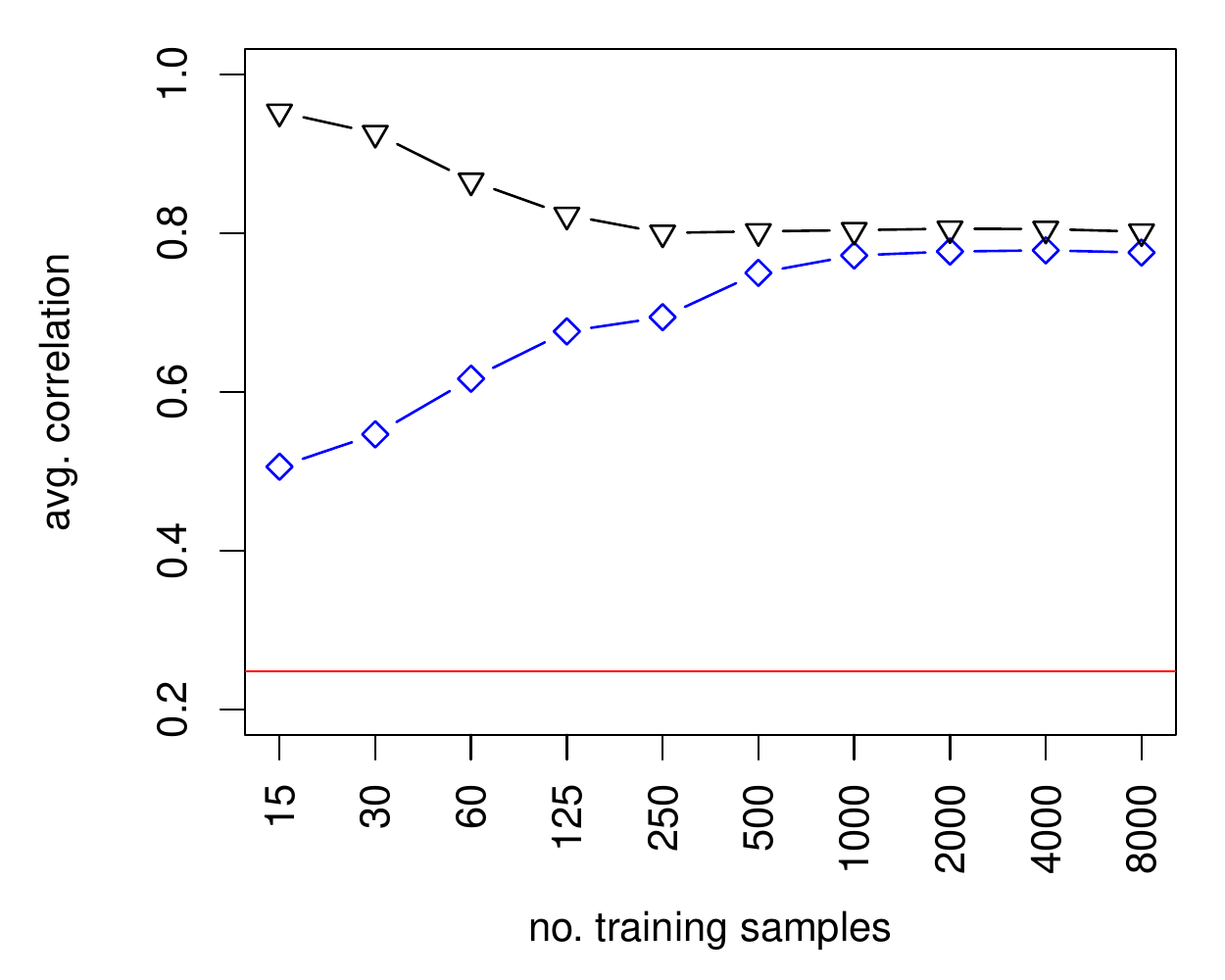}
\end{minipage}
\begin{minipage}{0.64\textwidth}
\caption{\label{perf_tr_size}
Performance with and without loss-function learning as a function of the size of the training set. Performance was assessed by calculating the average correlations between predicted and true cellular contributions over all cell types. The blue diamonds and black triangles correspond to the performance of loss-function learning for the validation mixtures and training mixtures, respectively. The performance of standard  DTD with $g=1$ is shown as a red line for the validation mixtures. }
\end{minipage}
\end{figure}

 We repeated the simulation in subsection \ref{T_cell_subtypes}, but varied the size of the training dataset. We observed that loss-function learning improved accuracy for 
training datasets as small as 15 samples. Moreover, with more training data added the boost in performance grew and saturated only for training sets with more than 1,000 samples 
(Figure \ref{perf_tr_size}).

\subsection{HPC-empowered loss-function learning rediscovers established cell markers and complements them by new discriminatory genes for improved performance}
\label{full_model}
Here, we introduce a final model, optimized on the 5,000 most variable genes. For this purpose, we generated 25,000 training mixtures from the melanomas of the training data. 
With standard desktop workstations the solution of this problem was computationally not feasible. A single computation of the gradient took 16 hours (2x Intel Xeon CPU [X5650; Nehalem Six Core, 2.67 GHz], 148 Gb RAM), and this needs to be computed several hundred times until convergence.
Therefore, we developed a High-Performance-Computing (HPC) implementation
of our code by parallelizing equations (\ref{L_out}) and (\ref{dCdg}) with MPI, using the
pbdMPI library (\cite{Chen2012pbdMPIpackage}, \cite{Chen2012pbdMPIvignette}) as an interface. Furthermore, we linked R with
the Intel Math Kernel Library for threaded and vectorized matrix
operations. We ran the algorithm on 25 nodes of our QPACE 3 machine \citep{Georg:2017zua} with 8 MPI tasks per node and 32 hardware threads per task, where each thread can use two AVX512 vector units.
In 16 hours 5,086 iterations were finished, after which the loss (\ref{L_out}) was stable to within 1\%.

\begin{figure}[h]
\centering
\includegraphics[width=1\textwidth]{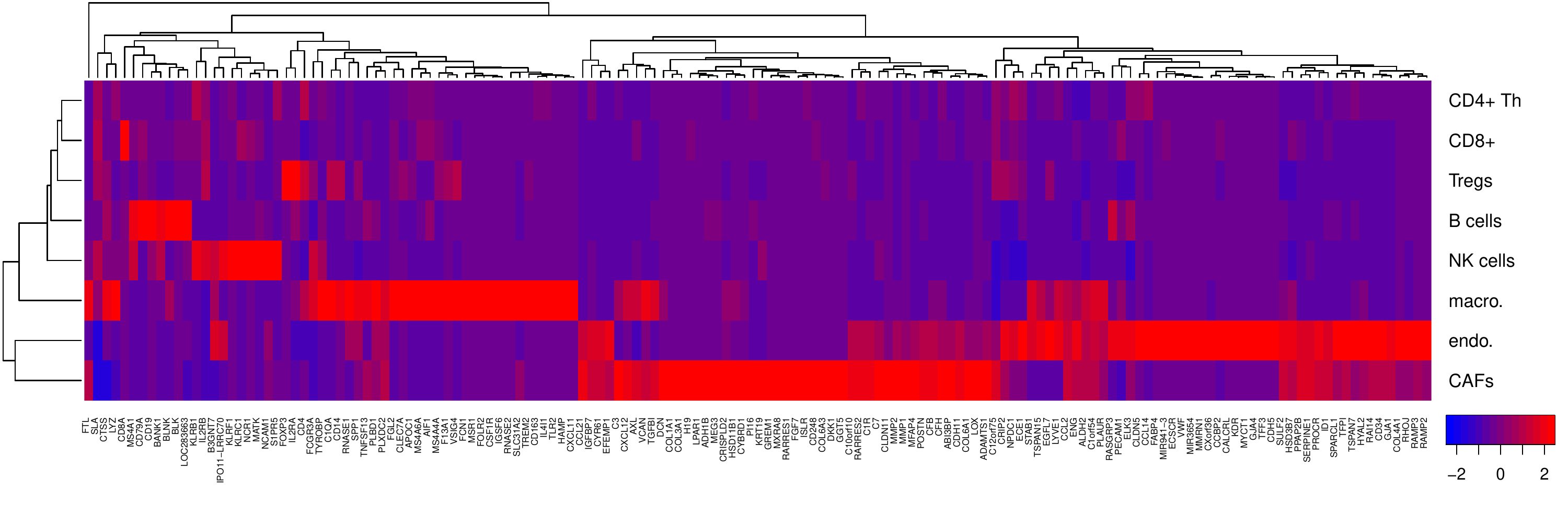}
\caption{\label{Heatmap_10000}Heatmap of $X$ for the features with the top 150 weights ($\hat g_i \times \textrm{var}(X_{i,\cdot})$). Blue corresponds to low expression and red to high expression. The data were clustered by Euclidean distance. }
\end{figure}

The high-performance model includes several genes, whose expression is characteristic for the cells distinguished in the present study. These include, among others, 
the CD8A gene, which encodes an integral membrane glycoprotein essential for the activation of cytotoxic T-lymphocytes \citep{VEILLETTE1988301} and the protection of 
a subset of NK cells against lysis, thus enabling them in contrast to CD8- NK cells to lyse multiple target cells \citep{IMM:IMM2235}. As evident from Figure \ref{Heatmap_10000}, 
NK cells are clearly set apart from all the other cell types studied by the expression of the killer cell lectin like receptor genes KLRB1, KLRC1, and KLRF1 \citep{doi:10.1146/annurev.immunol.19.1.197}. B cells, on the other hand, are clearly characterized by the expression of (i) CD19, which assembles with the antigen receptor of B lymphocytes and influences B-cell selection and differentiation \citep{RIS_0}, (ii) CD20 (MS4A1), which is coexpressed with CD19 and functions 
 as a store-operated calcium channel \citep{Li24102003}, (iii) B Lymphocyte Kinase (BLK), a src-family protein tyrosine kinase that plays an important role in B-cell receptor signaling and phosphorylates specifically (iv) CD79A at Tyr-188 and Tyr-199 as well as CD79B (not among the top 150 genes) at Tyr-196 and Tyr-207, which are required for the surface expression and function of the B-cell antigen receptor complex (\cite{HSUEH2000283}), and (v) BLNK,  which bridges BLK activation with downstream signaling pathways \citep{wienands1998slp}. 
The expression of FOXP3 is also highly cell specific. FOXP3 distinguishes regulatory T cells from other CD4+ cells  and functions as a master regulator of their 
 development and function (\cite{Hori1057}). Finally, CD4+ T-helper (Th) cells are distinguished indirectly from all the other aforementioned lymphocytes by the lack of expression of cell 
 type-specific genes. In contrast to lymphocytes, macrophages, cancer-associated fibroblasts (CAFs), and 
 endothelial cells, which line the interior surface of blood vessels and lymphatic vessels, are characterized each by a much larger number of
 genes. Exemplary genes include CD14, CD163, MSR1, STAB1, and CSF1R for macrophages. The monocyte differentiation antigen CD14, for instance, mediates the innate immune response 
 to bacterial lipopolysaccharide (LPS) by activating the NF-$\kappa$B pathway and cytokine secretion (\cite{HAZIOT1996407}), while the colony stimulating factor 1 receptor (CSF1R) acts as a receptor for the
 hematopoietic growth factor CSF1, which controls the proliferation and function of macrophages (\cite{SHERR1985665}). CAFs, on the other hand, are distinguished by the expression of genes encoding 
 extracellular matrix proteins such as fibulin-2 (FBLN2) and fibulin-3 (EFEMP1), various collagens (COL1A1, COL3A1, COL6A1, COL6A3), versican (VCAN), a well known mediator of cell-to-cell and cell-to-matrix interactions (\cite{Jiong}) that plays critical roles in 
 cancer biology (\cite{DUwil}), as well as the matrix metalloproteinases MMP1 and MMP2, two collagen degrading enzymes that allow cancer cells to migrate out of the primary
 tumor to form metastases (\cite{GUPTA201465}). 
 Noteworthy is also GREM1, an antagonist of the bone morphogenetic protein pathway. Its expression and secretion by stromal cells in tumor tissues promotes the survival and proliferation of cancer cells \citep{sneddon2006bone}. 
 Genes characteristic for endothelial cells include among others CDH5, a member of the cadherin superfamily essential for endothelial adherens junction 
 assembly and maintenance (\cite{Gory-Faure2093}), the endothelial cell-specific chemotaxis receptor (ECSCR) gene, which encodes a cell-surface single-transmembrane domain glycoprotein that plays a role in endothelial cell migration, apoptosis and proliferation (\cite{10.1371/journal.pone.0021482}), claudin-5 (CLDN5), which forms the backbone of tight junction strands between endothelial cells \citep{haseloff2015transmembrane}, and the von Willebrand factor (VWF), which mediates the adhesion of platelets to sites of vascular damage by binding to specific platelet membrane glycoproteins and to constituents of exposed connective tissue (\cite{doi:10.1146/annurev.biochem.67.1.395}).

We discussed 29 genes of the top 150 shown in Figure \ref{Heatmap_10000}. These genes have a total weight of 28\% of all 5,000 gene weights (calculated as $\hat g_i \times \textrm{var}(X_{i,\cdot})$). Our algorithm complements this gene set with additional genes, including many that were, to our knowledge, not yet used to characterize cell types. An interesting example is CXorf36 (DIA1R), which has been described as being expressed at low levels in many tissues and deletion and/or mutations of which have been associated with autism spectrum disorders \citep{aziz2011dia1r}. However, nothing is known about its function to date. Therefore, its observed overexpression in endothelial cells may provide an important clue for future study on its function.

\subsection{Loss-function learning shows similar performance as CIBERSORT for the dominating cell populations
and improves accuracy for small populations and in the distinction of closely related cell types} 

\begin{figure}
\centering
\includegraphics[width=.8\textwidth]{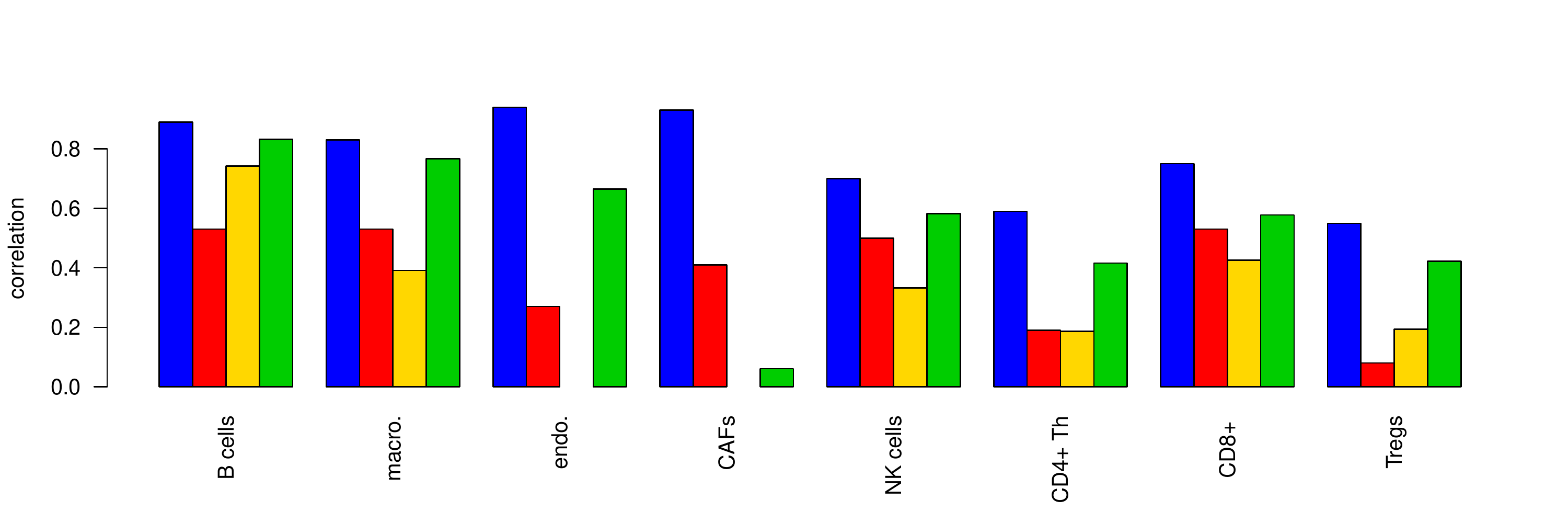}
\caption{\label{summary}Performance comparison. The methods are from left to right:  loss-function learning (blue), standard DTD with $g=1$ on the 5,000 most variable genes (red), CIBERSORT${}^{a}$ (yellow), CIBERSORT${}^{b}$ (green). Performance was calculated as correlation between predicted and true frequencies on 1,000 validation mixtures. Endothelial cells (endo.) and CAFs were not estimated by CIBERSORT$^{a}$ and thus are not shown.}
\end{figure}

Next we compared our model trained in subsection \ref{full_model} to a competing method. For this, we generated 1,000 test mixtures from our validation melanomas. We chose CIBERSORT \citep{newman2015robust} for comparison, because it was consistently among the 
best DTD algorithm in a broad comparison of five different algorithms on several benchmark datasets \citep{newman2015robust}. We ran CIBERSORT on the test mixtures, using two distinct approaches: first, we uploaded 
our validation data to CIBERSORT using their reference profiles. The performance is summarized in Figure \ref{summary} as CIBERSORT${}^a$ (yellow). We observed that the large population of B cells was estimated accurately, while smaller populations 
were inaccurate (NK cells, Tregs). Next, we uploaded our reference profiles and used the CIBERSORT gene selection (CIBERSORT${}^b$ green). We found that highly abundant cell types (B cells and CD8+ T cells) were predicted with high accuracy. However, 
the distinction of similar cell types such as CD4+ T helper cells and Tregs was compromised, $r=0.42$ and $r=0.42$, respectively. 
Similarly, predictions for the small populations of CAFs were compromised. That might be explained 
by the fact that CIBERSORT does not take into account their distinction and thus appropriate marker genes might be missing. 
In a direct comparison to CIBERSORT our method showed similar or better performance.

\begin{figure}
\centering
\includegraphics[width=1\textwidth]{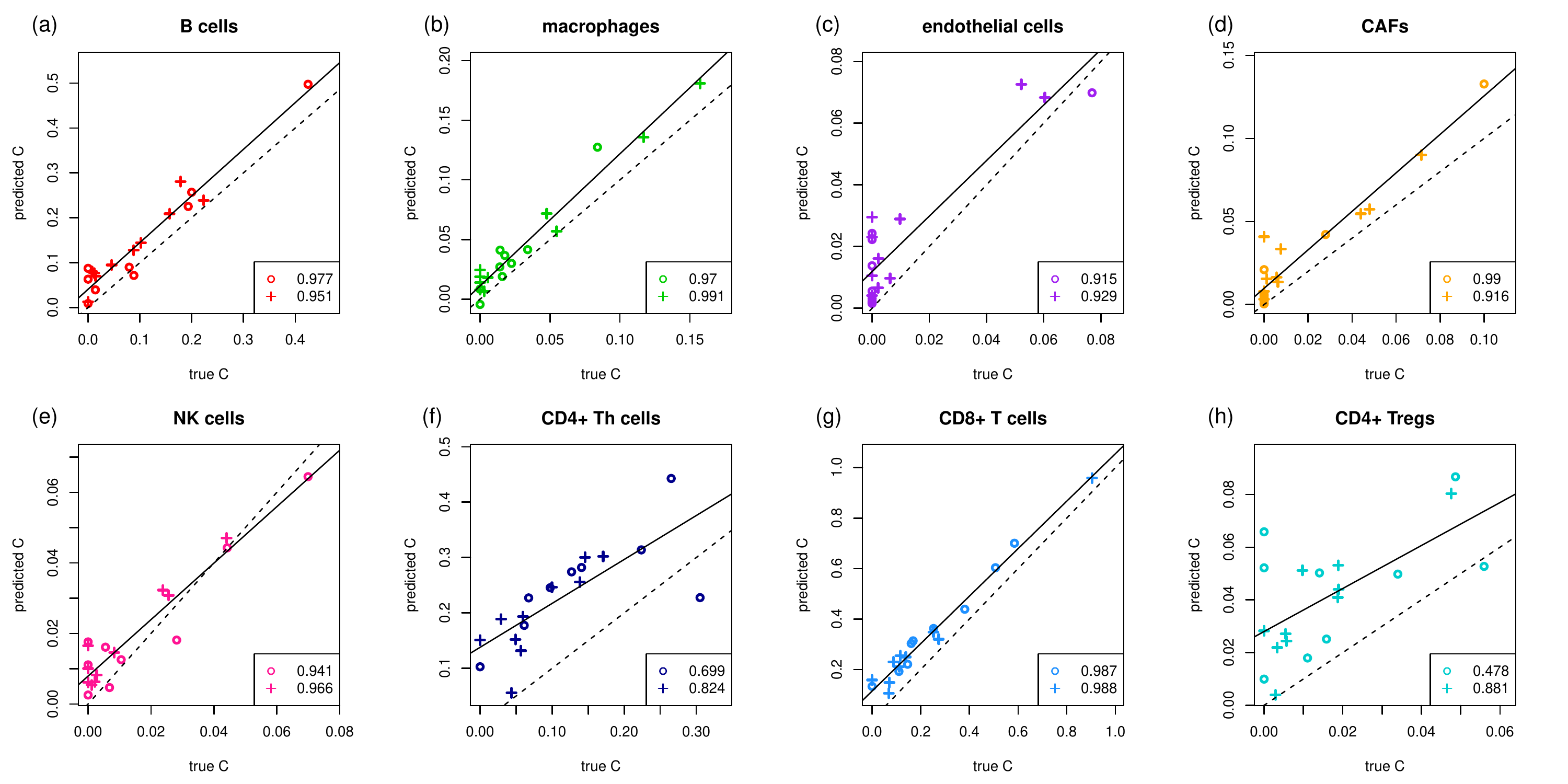}
\caption{\label{All_cells}Deconvolution of melanoma tissues. The circles indicate melanomas from the validation data and plusses from the training data. Figure (a) to (h) correspond to B cells, macrophages, endothelial cells, CAFs, NK cells, CD4+ Th cells, CD8+ T cells, and CD4+ Tregs, respectively. The solid black lines show the corresponding linear regression fits, the dashed lines the identity.  }
\end{figure}

\subsection{Loss-function learning improves the decomposition of bulk melanoma profiles}
All mixtures discussed so far were artificial because only 100 single-cell profiles were chosen randomly. They might differ significantly from mixtures in real tissue. Therefore, we generated 19 full bulk melanoma profiles by summing up the respective single-cell profiles. These should reflect bulk melanomas \citep{marinov2014single}. Our predictions are contrasted with the true proportions in Figure \ref{All_cells}. Only the predictions for Tregs were compromised with $r=0.48$, while the predictions for all other cell types were reliable with correlations ranging from $r=0.70$ (CD4+ Th) to $r=0.99$ (CAFs) on the validation melanomas.

\section{Discussion}

We suggest loss-function learning for digital tissue deconvolution to adapt the deconvolution
algorithm to the requirements of specific application domains. The concept is similar to an
embedded feature-selection approach in regression or classification problems. In both contexts
feature selection is directly linked to a prediction algorithm and not treated as an independent
preprocessing step.    

We described and tested a specific instance of loss-function learning using squared residuals
for $\mathcal{L}_g$. The concept is not limited to this type of inner loss function
and can also be used in combination with other loss functions such as those from
penalized least-squares regression \citep{altboum2014digital}, $l_1$ regression, 
or support vector regression \citep{newman2015robust}.
However, the least-squares loss function allowed us to state the outer optimization problem in a closed analytical form,
reducing computational burden.  

The outer loss function $L$ evaluates the fit of estimated and true cellular proportions
in the training samples. We chose the correlation of estimated versus true quantities across samples, and no 
absolute measure of deviation such as $||c-\hat c||_2^2$. Moreover, we did not require the estimated 
proportions $\hat C_{\cdot,k}$ for tissue $k$ to sum up to one.
Consequently, when testing our method we did not look at absolute deviations of true versus estimated cell proportions
but only at their correlation. We cannot reliably infer how many cells of a specific type (e.g., T cells)
are in a tissue (Figure \ref{All_cells}), nor can we infer whether they constituted 10\% or 20\% of the cells in this tissue.
However, if we had two tissues and estimated that there were more cells of that type in the first tissue compared to the second, this relation was also found in the true cell populations.
Calibrating the method to absolute cell numbers or percentages within a tissue appears to be difficult.
This might be due to the normalization of the libraries from single cells to a common size, which might not reflect
true biology if, for example, some cell types contain more RNA than others. Post-hoc normalization of the estimated quantities might be an option.

In summary, we introduced loss-function learning as a new machine-learning approach to the digital tissue deconvolution
problem. It allows us to adapt to application-specific requirements such as focusing on small cell populations or delineating similar cell types.
In simulations and in an application to melanoma tissues our method quantified large cell fractions as
accurately as existing methods and significantly improved the detection of small cell populations and the distinction of similar cell types.
 
\section{Acknowledgement}
This work was supported by BMBF (eMed Grant 031A428A)  and DFG (FOR-2127  and SFB/TRR-55).

\bibliography{Altenbuchinger}
\bibliographystyle{apalike}
\end{document}